\documentclass[a4paper,11pt]{article}
\usepackage{pos}
\newcommand{\nn}{\nonumber}
\newcommand{\be}{\begin{equation}}
\newcommand{\ee}{\end{equation}}
\newcommand{\bea}{\begin{eqnarray}}
\newcommand{\eea}{\end{eqnarray}}
\newcommand{\spur}[1]{\not\! #1 \,}
\newcommand{\dd}{\displaystyle}
\title{Probing New Physics with heavy hadron decays}
\author[a]{Pietro Colangelo}
\author*[a]{Fulvia De Fazio}
\author[a,b]{Francesco Loparco}

\affiliation[a]{Istituto Nazionale di Fisica Nucleare, Sezione di Bari\\Via Orabona 4,
I-70126 Bari, Italy}

\affiliation[b]{Universit\'a degli Studi  di Bari\\ Via Orabona 4,
I-70126 Bari, Italy}

\emailAdd{pietro.colangelo@ba.infn.it}
\emailAdd{fulvia.defazio@ba.infn.it}
\emailAdd{francesco.loparco1@ba.infn.it}

\abstract{The increasing number of flavour anomalies calls for the investigation of new processes where tensions similar to the observed ones  could emerge. Observables  sensitive to physics beyond the Standard Model need to  be identified. We discuss the inclusive semileptonic decays of polarized beauty baryons, computed  through the heavy quark expansion  at  ${\cal O}(1/m_b^3)$ and  at the leading order in $\alpha_s$.  We account for New Physics interactions in a model-independent way,  extending the  Standard Model  $b \to U \ell {\bar \nu}_ \ell$ low energy Hamiltonian  (with $U=u,\,c$ and $\ell=e,\,\mu,\,\tau$) with the inclusion of the full  set of D=6 semileptonic operators with left-handed neutrinos. We identify a set of promising observables, the study of which can  be included in the physics programmes of  future facilities, such as FCC-e.}

\FullConference{%
  *** The European Physical Society Conference on High Energy Physics (EPS-HEP2021), ***\\
  *** 26-30 July 2021 ***\\
  *** Online conference, jointly organized by Universität Hamburg and the research center DESY ***
}

\begin{document}
\maketitle

\section{Introduction}
In the searches for signals of New Physics (NP), deviations in a number of observables with respect to the Standard Model (SM) predictions have recently been detected. They constitute the so-called {\it flavour anomalies}, and  show up  in selected   tree-level and loop-induced decays of  $B$, $B_s$ and $B_c$ mesons \cite{Alguero:2021anc}. In particular,  hints of lepton flavour universality (LFU) violations have been collected. The observed deviations call for new investigations, in particular focused on other  heavy hadron decay processes. Here, we analyze the inclusive semileptonic beauty baryon decays \cite{Colangelo:2020vhu},   for which the nonperturbative effects of strong interactions can be systematically treated exploiting an expansion in the inverse heavy quark mass \cite{Chay:1990da,Bigi:1993fe}. 
 
Even though 
the formalism holds for a generic baryon comprising a heavy quark, we focus on inclusive $\Lambda_b\to  X_{c,u} \ell^- {\bar \nu}_\ell$ decays, performing the  heavy quark mass expansion (HQE)   at  ${\cal O}(1/m_Q^3)$. At each order in the expansion,
  a number of hadronic matrix elements is required; they are parametrized in terms of basic non perturbative quantities. Our main result is the derivation of 
 the  baryon matrix elements at  ${\cal O}(1/m_Q^3)$  in the case of a polarized baryon. 
 The result is applied to derive the fully differential decay rate for $\Lambda_b\to  X_{c,u} \ell^- {\bar \nu}_\ell$ both in 
 the SM  and in an extension of the SM effective weak Hamiltonian  comprising  vector, scalar, pseudoscalar, tensor and axial  operators. Other studies extending the effective Hamiltonian in a similar way are in   \cite{Biancofiore:2013ki,Colangelo:2018cnj,Bhattacharya:2018kig,Colangelo:2016ymy,Mannel:2017jfk,Kamali:2018fhr,Kamali:2018bdp}.
 
At  LHC the  $\Lambda_b$ is  produced unpolarized  \cite{Aaij:2013oxa,Aad:2014iba,Sirunyan:2018bfd,Aaij:2020vzk} since  the $b$ quark   mainly comes from QCD processes.  A sizable longitudinal $\Lambda_b$ polarization is expected for $b$ quarks  produced in $Z$ and top quark decays   \cite{BUSKULIC1996437,Abbiendi:1998uz,2000205}.   The investigations  of   effects beyond the Standard Model (BSM) in the polarized case must  be optimized in such an experimental environment. Our analysis and results 
are presented below. 

\section{Generalized effective weak Hamiltonian}\label{hamiltonian}
We consider  a  beauty baryon $H_b$   with  spin $s$.  The inclusive semileptonic 
$H_b(p,s) \to X_U(p_X) \ell^-(p_\ell) {\bar \nu_\ell}(p_\nu) $ decays ($U=u,\,c$), induced by the $b \to u,\,c$ transitions,
are   described by  the general low-energy  Hamiltonian,  which extends the SM one:
\bea
&& \hskip -.4cm
H_{\rm eff}^{b \to U \ell \nu}= \frac{G_F}{\sqrt{2}} V_{Ub} \label{hamil} \\
&&  \hskip -.4cm\Big[(1+\epsilon_V^\ell) \left({\bar U} \gamma_\mu (1-\gamma_5) b \right)\left( {\bar \ell} \gamma^\mu (1-\gamma_5) {\nu}_\ell \right)
+ \epsilon_S^\ell \, ({\bar U} b) \left( {\bar \ell} (1-\gamma_5) { \nu}_\ell \right)
+ \epsilon_P^\ell \, \left({\bar U} \gamma_5 b\right)  \left({\bar \ell} (1-\gamma_5) { \nu}_\ell \right) \nn \\
&& \hskip -.4cm+ \epsilon_T^\ell \, \left({\bar U} \sigma_{\mu \nu} (1-\gamma_5) b\right) \,\left( {\bar \ell} \sigma^{\mu \nu} (1-\gamma_5) { \nu}_\ell \right) 
+  \epsilon_R^\ell \left({\bar U} \gamma_\mu (1+\gamma_5) b \right)\left( {\bar \ell} \gamma^\mu (1-\gamma_5) {\nu}_\ell \right)   \Big] + h.c.\,\,\,  . \nn
\eea
 $\epsilon^\ell_{V,S,P,T,R}$ are complex  and lepton-flavour dependent coefficients. Only left-handed  neutrinos are considered,  and $V_{Ub}$ is the relevant Cabibbo-Kobayashi-Maskawa (CKM) matrix element.   
The Hamiltonian   can be written as
\be
H_{\rm eff}^{b \to U \ell \nu}= \frac{G_F}{\sqrt{2}} V_{Ub} \sum_{i=1}^5 C_i^\ell \, J^{(i)}_M\, L^{(i)M} + h.c.\,\,\, ,\label{hnew}\ee
with  $C_1^\ell=(1 +\epsilon_V^\ell)$ and  $C_{2,3,4,5}^\ell=\epsilon_{S,P,T,R}^\ell$.   $J_M^{(i)}$ ($L^{(i)M}$) is the hadronic (leptonic) current in each operator,  $M$ are set of Lorentz indices contracted between $J$ and $L$. The SM expression is recovered for $i=1$ and $\epsilon^\ell_{V,S,P,T,R}=0$. We keep  $m_\ell \neq 0$ for all leptons $\ell=e,\mu,\tau$.
\section{Inclusive decay width}\label{OPE}
The  $H_b$  inclusive semileptonic differential  decay width  can be written as
\be
d\Gamma= d\Sigma \, \frac{G_F^2 |V_{Ub}|^2}{4m_H} \sum_{i,j} C_i^* C_j (W^{ij})_{MN} (L^{ij})^{MN} .
\ee
 $G_F$ is the Fermi constant, $q=p_\ell+p_\nu$  is the lepton pair momentum, and   $d\Sigma$  is the phase-space element  $d\Sigma=(2\pi) d^4q \, \delta^4(q-p_\ell-p_\nu) [dp_\ell]\,[dp_\nu] $, with   $[dp]=\displaystyle\frac{d^3 p}{(2\pi)^3 2p^0}$. The leptonic tensor is  $(L^{ij})^{MN}=  L^{(i)\dagger M} L^{(j)N}$.
On the basis of the optical theorem, the hadronic tensor  $(W^{ij})_{MN}=\frac{1}{\pi}{\rm Im}(T^{ij})_{MN}$ is given in terms  of the forward  amplitude
\bea
(T^{ij})_{MN}&=&i\,\int d^4x \, e^{-i\,q \cdot x} \langle H_b(p,s)|T[ J^{(i)\dagger}_M (x) \,J^{(j)}_N (0)] |H_b(p,s) \rangle\,\,.\label{Tij-gen}
\eea
 The hadron momentum $p=m_b v+k$ is written in terms of the  four-velocity $v$,  the heavy quark mass $m_b$ and  a residual momentum $k$ of order $\Lambda_{QCD}$. The QCD $b$ quark field  $b(x)$ is redefined as  $b(x)=e^{-i\,m_b v \cdot x} b_v(x)$, with   
 $b_v(x) $ satisfying the equation
$b_v(x)=\dd{\left(P_+ +\frac{i {\spur D}}{2m_b} \right)} b_v(x)$,
with $P_+=\displaystyle\frac{1+ \spur v}{2}$  the velocity projector. Introducing $p_X=m_bv+k-q$ we have:
\be
(T^{ij})_{MN}=\langle H_b(v,s)|{\bar b}_v(0) \Gamma_M^{(i)\dagger} S_U(p_X) \Gamma_N^{(j)}b_v(0) |H_b(v,s)\rangle\,\, ,
\ee
with  $S_U(p_X)$  the $U$ quark propagator. The heavy quark expansion  in powers of  $m_b^{-1}$  is carried out \cite{Chay:1990da,Bigi:1993fe}. It is obtained replacing $k \to iD$ ($D$ is  the QCD covariant derivative) and expanding
$S_U(p_X)=S_U^{(0)}-S_U^{(0)}(i {\spur D})S_U^{(0)}+S_U^{(0)}(i {\spur D})S_U^{(0)}(i {\spur D})S_U^{(0)}+ \dots\,\,$
where $S_U^{(0)}=\displaystyle\frac{1}{m_b {\spur v}-{\spur q} -m_U}$.
Writing $p_U=m_b v -q$, ${\cal P}=({\spur p}_U+m_U)$ and $\Delta_0=p_U^2-m_U^2$,  the expansion  at order $1/m_b^3$ reads:
\bea
\frac{1}{\pi}{\rm Im}(T^{ij})_{MN} \hskip-0.3cm&&\hskip -.3cm= \frac{1}{\pi}{\rm Im}\frac{1}{\Delta_0}\langle H_b(v,s)|{\bar b}_v [\Gamma_M^{(i)\dagger} {\cal P} \Gamma_N^{(j)}]b_v |H_b(v,s) \rangle +\nn \\
\hskip -.9cm&& \hskip -1cm-\frac{1}{\pi}{\rm Im}\frac{1}{\Delta_0^2}\langle H_b(v,s)|{\bar b}_v[ \Gamma_M^{(i)\dagger} {\cal P}\gamma^{\mu_1}{\cal P} \Gamma_N^{(j)}](i D_{\mu_1})b_v |H_b(v,s)\rangle +\label{expansion} \\
\hskip -.9cm&& \hskip -1cm +\frac{1}{\pi}{\rm Im}\frac{1}{\Delta_0^3}\langle H_b(v,s)|{\bar b}_v[ \Gamma_M^{(i)\dagger} {\cal P}\gamma^{\mu_1}
{\cal P}\gamma^{\mu_2}{\cal P} \Gamma_N^{(j)}](i D_{\mu_1})(i D_{\mu_2})b_v |H_b(v,s)\rangle +\nn
\\
\hskip -.9cm&& \hskip -1cm -\frac{1}{\pi}{\rm Im}\frac{1}{\Delta_0^4} \langle H_b(v,s)|{\bar b}_v[ \Gamma_M^{(i)\dagger} {\cal P}\gamma^{\mu_1}
{\cal P}\gamma^{\mu_2}{\cal P} \gamma^{\mu_3}{\cal P}\Gamma_N^{(j)}](i D_{\mu_1})(i D_{\mu_2})(i D_{\mu_3})b_v |H_b(v,s) \rangle .\nn 
\eea
The expression \eqref{expansion} involves  the hadronic matrix elements 
\be
{\cal M}_{\mu_1 \dots \mu_n}=\langle H_b(v,s)|({\bar b}_v)_a(i D_{\mu_1})\dots(i D_{\mu_n})(b_v)_b |H_b(v,s)\rangle , \label{matel}
\ee  
with $a,b$  Dirac indices.  The matrix elements 
can be written  in terms of  nonperturbative parameters, the number of which  increases with the order of the expansion.
At ${\cal O}(1/m_b^3)$  the required ones are:
\bea
\langle H_b(v,s)|{\bar b}_v (iD)^2 b_v|H_b(v,s)\rangle&=&-2m_H\,{\hat \mu}_\pi^2 \label{mupi}\\
\langle H_b(v,s)|{\bar b}_v (iD_\mu)(iD_\nu)(-i \sigma^{\mu \nu})b_v|H_b(v,s)\rangle&=&2m_H\,{\hat \mu}_G^2 \label{mug} 
\\
\langle H_b(v,s)|{\bar b}_v (iD_\mu)(i v \cdot D) (i D^\mu) b_v|H_b(v,s)\rangle&=&2m_H\,{\hat \rho}_D^3\label{rd} \\
\langle H_b(v,s)|{\bar b}_v (iD_\mu)(i v \cdot D) (i D_\nu) (-i \sigma^{\mu \nu})b_v|H_b(v,s)\rangle &=&2m_H\,{\hat \rho}_{LS}^3 \,\, .\label{rls}
 \eea
A method to compute  ${\cal M}_{\mu_1 \dots \mu_n}$ is proposed in \cite{Dassinger:2006md,Mannel:2010wj}. In the case of  baryons  the dependence on the spin $s_\mu$  in \eqref{matel} must be kept.  In \cite{Colangelo:2020vhu} 
${\cal M}_{\mu_1 \dots \mu_n}$ have been  derived  at order $1/m_b^3$ for a polarized baryon,  considering all  operators in (\ref{hamil}). This extends previous results
 in \cite{Kamali:2018bdp,Mannel:2017jfk,Grossman:1994ax,Manohar:1993qn,Balk:1997fg}. 
 
From the expressions of the matrix elements ${\cal M}_{\mu_1 \dots \mu_n}$  the hadronic tensor  can be computed and  expanded  in Lorentz structures which depend on $v$, $q$ and $s$. The results  for the SM and for  the effective Hamiltonian Eq.~\eqref{hamil} are collected  in   \cite{Colangelo:2020vhu}. 
The four-fold differential decay rate for the  $H_b(v,s) \to X(p_X) \ell^-(p_\ell) {\bar \nu}_\ell(p_\nu)$ transition  reads
\be
\frac{d^4 \Gamma}{dq^2 \,d(v \cdot q) \,dE_\ell  \,d \cos \theta_P}=\frac{G_F^2 |V_{Ub}|^2}{32 (2 \pi)^3 m_H} \sum_{i,j}C_i^* C_j \frac{1}{\pi}{\rm Im}(T^{ij})_{MN}(L^{ij})^{MN} \,\, ,\label{full}
\ee
with  $p_\ell=(E_\ell,{\vec p}_\ell)$.  $\theta_P$   the angle between ${\vec p}_\ell$ and ${\vec s}$  in the $H_b$ rest frame.
Double and single  decay distributions are obtained  integrating (\ref{full}) over the phase-space \cite{Jezabek:1996ia}.  Performing all  integrations, the full decay width  can be  written as:
\be
\Gamma(H_b \to X \ell^- {\bar \nu}_\ell)=\Gamma_b\sum_i  \left\{C_0^{(i)}+\frac{\mu^2_\pi}{m_b^2} C_{\mu_\pi^2}^{(i)}+\frac{\mu^2_\pi}{m_b^2} C_{\mu_G^2}^{(i)}+\frac{\rho_D^3}{m_b^3} C_{\rho_D^3}^{(i)}+\frac{\rho_{LS}^3}{m_b^3} C_{\rho_{LS}^3}^{(i)} \right\} , \label{fullwidth}
\ee
with $\Gamma_b=\displaystyle\frac{G_F^2 m_b^5 V_{Ub}^2}{192 \pi^3}$. The index  $i$ runs over the contribution of the various operators and of their interferences.
The   coefficients  $C^{(i)}$ depend on the NP couplings    in (\ref{hamil}) and  can be found in \cite{Colangelo:2020vhu}.

 The OPE breaks down in the endpoint region of the spectra, where   singularities  appear.  They require to be resummed in a $H_b$   shape function, and  the convolution  with such a function  smears the spectra at the endpoint. We have not considered the effects of the baryon shape function, which  is not known at present.
Perturbative QCD corrections are also not included:  in the SM various corrections have been computed
\cite{Czarnecki:1994bn,Jezabek:1996db,DeFazio:1999ptt,Trott:2004xc,Aquila:2005hq,Alberti:2014yda}.

\section{Results for $\Lambda_b \to X_{c,u} \ell \bar \nu_\ell$}\label{results}
We collect in this Section some  results obtained  for  observables in $\Lambda_b\to X_{u,c} \ell^- {\bar \nu}_\ell$. We refer to  \cite{Colangelo:2020vhu}  for the input parameters. For the  couplings  $\epsilon^\ell_{V,S,P,T,R}$ in  (\ref{hamil}),
for $U=u$ we use the  ranges  fixed   in  \cite{Colangelo:2019axi}. 
For $U=c$ we fix three NP benchmark points, set in  \cite{Colangelo:2018cnj} and \cite{Shi:2019gxi}.

In  the SM,
using $|V_{cb}|=0.042$, $|V_{ub}|=0.0037$ and  $\tau_{\Lambda_b}=(1.471\pm 0.009) \times 10^{-12}$ s \cite{Zyla:2020}, we obtain:
${\cal B}(\Lambda_b \to X_c  \mu {\bar \nu}_\mu)=11.0 \times 10^{-2}$,
${\cal B}(\Lambda_b \to X_c  \tau {\bar \nu}_\tau)=2.4 \times 10^{-2}$,
${\cal B}(\Lambda_b \to X_u  \mu {\bar \nu}_\mu)=11.65 \times 10^{-4}$ and
${\cal B}(\Lambda_b \to X_u  \tau {\bar \nu}_\tau)=2.75 \times 10^{-4}$.
For comparison, the available measurements are
${\cal B}(\Lambda_b \to \Lambda_c \ell^- {\bar \nu}_\ell + {\rm anything} )=(10.9 \pm 2.2)\times 10^{-2}$ 
 $\ell=e,\,\mu$,  and 
 ${\cal B}(\Lambda_b \to p \mu^- {\bar \nu})=(4.1 \pm 1.0) \times 10^{-4} $   \cite{Zyla:2020}.

\begin{figure}[t!]
\begin{center}
\includegraphics[width = 0.5\textwidth]{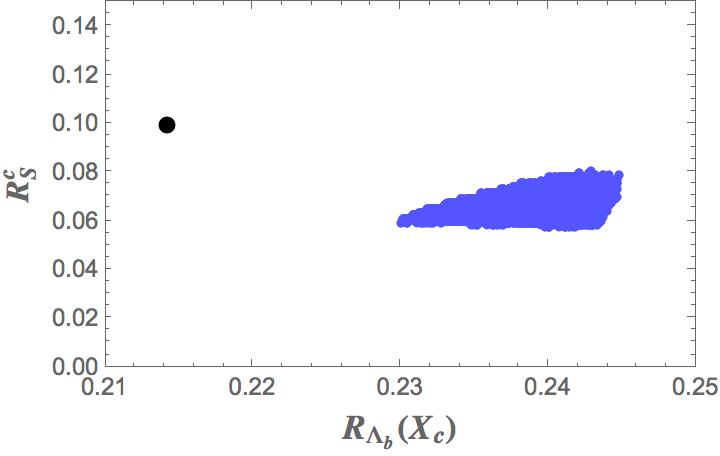} 
\caption{ \baselineskip 12pt  \small
 correlation between $R_{\Lambda_b}(X_c)$ and  $R_S^c$. The dot corresponds to SM, the broad region to NP.}\label{fig:lpslope}
\end{center}
\end{figure}

Among the various interesting observables, for  polarized $\Lambda_b$  we mention here the distribution 
$\dd \frac{d \Gamma (\Lambda_b \to X_U \ell {\bar \nu}_\ell)}{d\cos \theta_P}=A_\ell^U+B_\ell^U \, \cos \theta_P$. In  the ratio   $R_{\Lambda_b}(X_U)=\dd{\frac{\Gamma(\Lambda_b \to X_U \, \tau \, {\bar \nu}_\tau)}{\Gamma(\Lambda_b \to X_U \, \mu \, {\bar \nu}_\mu)} }=\dd \frac{A_\tau^U}{A_\mu^U}$  several theoretical uncertainties cancel out. 
The predictions in SM and in NP at the chosen benchmark points are: 
$R_{\Lambda_b}(X_u)^{SM} = 0.234$, $R_{\Lambda_b}(X_u)^{NP}  = 0.238$, $
R_{\Lambda_b}(X_c)^{SM} = 0.214 $, $R_{\Lambda_b}(X_c)^{NP} = 0.240$,
  at the leading order in $\alpha_s$. 
Also the ratio $R_S^U=B_\tau^U/B_\mu^U$ is  sensitive to  NP.  In  SM we find: $R_S^c=0.1$ and $R_S^u=0.08$. To study the   correlation between $R_{\Lambda_b}(X_U)$ and  $R_S^U$ we consider,
as an example, the generalized effective Hamiltonian  extended  only with the  tensor operator.   The  correlation plot 
 in  Fig.~\ref{fig:lpslope} shows  that, although experimentally challenging,  the measurement of  the two ratios discriminates  SM from NP.
\section{Conclusions}
We have described  the calculation of the inclusive semileptonic decay width of a polarized heavy hadron  at order ${\cal O}(1/m_b^3)$ in the HQE, at leading order in $\alpha_s$ and for non vanishing charged lepton mass. NP is considered  extending
 the SM effective Hamiltonian  including the full set of $D=6$ semileptonic operators. Among the  various result, the correlation   in  Fig.~\ref{fig:lpslope} shows the discriminating power between SM and NP which can be obtained by the analysis of inclusive polarized $\Lambda_b$ semileptonic modes.

\acknowledgments 
This study has been  carried out within the INFN project (Iniziativa Specifica) QFT-HEP.

\bibliographystyle{JHEP}
\bibliography{refsFFP2}

\providecommand{\href}[2]{#2}\begingroup\raggedright\begin{thebibliography}{10}

\bibitem{Alguero:2021anc}
M.~Alguer\'o et~al., {\it {$\boldsymbol{b\to s\ell\ell}$ global fits after
  Moriond 2021 results}},  in {\em {55th Rencontres de Moriond on QCD and High
  Energy Interactions}}, 4, 2021.
\newblock \href{http://arxiv.org/abs/2104.08921}{{\tt arXiv:2104.08921}}.

\bibitem{Colangelo:2020vhu}
P.~Colangelo, F.~De~Fazio, and F.~Loparco, {\it {Inclusive semileptonic
  $\Lambda_{b}$ decays in the Standard Model and beyond}},  {\em JHEP} {\bf 11}
  (2020) 032, [\href{http://arxiv.org/abs/2006.13759}{{\tt arXiv:2006.13759}}].

\bibitem{Chay:1990da}
J.~Chay, H.~Georgi, and B.~Grinstein, {\it {Lepton energy distributions in
  heavy meson decays from QCD}},  {\em Phys. Lett. B} {\bf 247} (1990)
  399--405.

\bibitem{Bigi:1993fe}
I.~I. Bigi et~al., {\it {QCD predictions for lepton spectra in inclusive heavy
  flavor decays}},  {\em Phys. Rev. Lett.} {\bf 71} (1993) 496--499,
  [\href{http://arxiv.org/abs/hep-ph/9304225}{{\tt hep-ph/9304225}}].

\bibitem{Biancofiore:2013ki}
P.~Biancofiore, P.~Colangelo, and F.~De~Fazio, {\it {On the anomalous
  enhancement observed in $B \to D^{(*)}\tau{\bar \nu}_\tau$ decays}},  {\em
  Phys. Rev.} {\bf D87} (2013) 074010,
  [\href{http://arxiv.org/abs/1302.1042}{{\tt arXiv:1302.1042}}].

\bibitem{Colangelo:2018cnj}
P.~Colangelo and F.~De~Fazio, {\it {Scrutinizing $ \overline{B}\to
  {D}^{\ast}\left(D\pi \right){\ell}^{-}{\overline{\nu}}_{\ell } $ and $
  \overline{B}\to {D}^{\ast}\left(D\gamma
  \right){\ell}^{-}{\overline{\nu}}_{\ell } $ in search of new physics
  footprints}},  {\em JHEP} {\bf 06} (2018) 082,
  [\href{http://arxiv.org/abs/1801.10468}{{\tt arXiv:1801.10468}}].

\bibitem{Bhattacharya:2018kig}
S.~Bhattacharya et~al., {\it {$b \rightarrow c \tau \nu _{\tau }$ Decays: a
  catalogue to compare, constrain, and correlate new physics effects}},  {\em
  Eur. Phys. J. C} {\bf 79} (2019) 268,
  [\href{http://arxiv.org/abs/1805.08222}{{\tt arXiv:1805.08222}}].

\bibitem{Colangelo:2016ymy}
P.~Colangelo and F.~De~Fazio, {\it {Tension in the inclusive versus exclusive
  determinations of $|V_{cb}|$: a possible role of new physics}},  {\em Phys.
  Rev.} {\bf D95} (2017) 011701, [\href{http://arxiv.org/abs/1611.07387}{{\tt
  arXiv:1611.07387}}].

\bibitem{Mannel:2017jfk}
T.~Mannel, A.~V. Rusov, and F.~Shahriaran, {\it {Inclusive semitauonic $B$
  decays to order ${\cal O} (\Lambda_{QCD}^3/m_b^3)$}},  {\em Nucl. Phys. B}
  {\bf 921} (2017) 211--224, [\href{http://arxiv.org/abs/1702.01089}{{\tt
  arXiv:1702.01089}}].

\bibitem{Kamali:2018fhr}
S.~Kamali, A.~Rashed, and A.~Datta, {\it {New physics in inclusive $B \to
  X_c\ell \bar{\nu}$ decay in light of $R(D^{(*)})$ measurements}},  {\em Phys.
  Rev. D} {\bf 97} (2018) 095034, [\href{http://arxiv.org/abs/1801.08259}{{\tt
  arXiv:1801.08259}}].

\bibitem{Kamali:2018bdp}
S.~Kamali, {\it {New physics in inclusive semileptonic $B$ decays including
  nonperturbative corrections}},  {\em Int. J. Mod. Phys.} {\bf A34} (2019)
  1950036, [\href{http://arxiv.org/abs/1811.07393}{{\tt arXiv:1811.07393}}].

\bibitem{Aaij:2013oxa}
{\bf LHCb} Collaboration, R.~Aaij et~al., {\it {Measurements of the
  $\Lambda_b^0 \to J/\psi \Lambda$ decay amplitudes and the $\Lambda_b^0$
  polarisation in $pp$ collisions at $\sqrt{s} = 7$ TeV}},  {\em Phys. Lett. B}
  {\bf 724} (2013) 27--35, [\href{http://arxiv.org/abs/1302.5578}{{\tt
  arXiv:1302.5578}}].

\bibitem{Aad:2014iba}
{\bf ATLAS} Collaboration, G.~Aad et~al., {\it {Measurement of the
  parity-violating asymmetry parameter $\alpha_b$ and the helicity amplitudes
  for the decay $\Lambda_b^0\to J/\psi+\Lambda^0$ with the ATLAS detector}},
  {\em Phys. Rev. D} {\bf 89} (2014) 092009,
  [\href{http://arxiv.org/abs/1404.1071}{{\tt arXiv:1404.1071}}].

\bibitem{Sirunyan:2018bfd}
{\bf CMS} Collaboration, A.~M. Sirunyan et~al., {\it {Measurement of the
  $\Lambda_b$ polarization and angular parameters in $\Lambda_b\to J/\psi\,
  \Lambda$ decays from pp collisions at $\sqrt{s}=$ 7 and 8 TeV}},  {\em Phys.
  Rev. D} {\bf 97} (2018) 072010, [\href{http://arxiv.org/abs/1802.04867}{{\tt
  arXiv:1802.04867}}].

\bibitem{Aaij:2020vzk}
{\bf LHCb} Collaboration, R.~Aaij et~al., {\it {Measurement of the
  $\Lambda^0_b\rightarrow J/\psi\Lambda$ angular distribution and the
  $\Lambda^0_b$ polarisation in $pp$ collisions}},
  \href{http://arxiv.org/abs/2004.10563}{{\tt arXiv:2004.10563}}.

\bibitem{BUSKULIC1996437}
{\bf ALEPH} Collaboration, D.~Buskulic et~al., {\it {Measurement of the
  $\Lambda_b$ polarization in Z decays}},  {\em Phys. Lett. B} {\bf 365} (1996)
  437 -- 447.

\bibitem{Abbiendi:1998uz}
{\bf OPAL} Collaboration, G.~Abbiendi et~al., {\it {Measurement of the average
  polarization of b baryons in hadronic $Z^0$ decays}},  {\em Phys. Lett. B}
  {\bf 444} (1998) 539--554, [\href{http://arxiv.org/abs/hep-ex/9808006}{{\tt
  hep-ex/9808006}}].

\bibitem{2000205}
{\bf DELPHI} Collaboration, P.~Abreu et~al., {\it {$\Lambda_b$ polarization in
  $Z^0$ decays at LEP}},  {\em Phys. Lett. B} {\bf 474} (2000) 205 -- 222.

\bibitem{Dassinger:2006md}
B.~M. Dassinger, T.~Mannel, and S.~Turczyk, {\it {Inclusive semi-leptonic B
  decays to order 1 / m(b)**4}},  {\em JHEP} {\bf 03} (2007) 087,
  [\href{http://arxiv.org/abs/hep-ph/0611168}{{\tt hep-ph/0611168}}].

\bibitem{Mannel:2010wj}
T.~Mannel, S.~Turczyk, and N.~Uraltsev, {\it {Higher Order Power Corrections in
  Inclusive B Decays}},  {\em JHEP} {\bf 11} (2010) 109,
  [\href{http://arxiv.org/abs/1009.4622}{{\tt arXiv:1009.4622}}].

\bibitem{Grossman:1994ax}
Y.~Grossman and Z.~Ligeti, {\it {The Inclusive $\bar B \to \tau \bar \nu X$
  decay in two Higgs doublet models}},  {\em Phys. Lett. B} {\bf 332} (1994)
  373--380, [\href{http://arxiv.org/abs/hep-ph/9403376}{{\tt hep-ph/9403376}}].

\bibitem{Manohar:1993qn}
A.~V. Manohar and M.~B. Wise, {\it {Inclusive semileptonic B and polarized
  Lambda(b) decays from QCD}},  {\em Phys. Rev.} {\bf D49} (1994) 1310--1329,
  [\href{http://arxiv.org/abs/hep-ph/9308246}{{\tt hep-ph/9308246}}].

\bibitem{Balk:1997fg}
S.~Balk, J.~Korner, and D.~Pirjol, {\it {Inclusive semileptonic decays of
  polarized $\Lambda_b$ baryons into polarized $\tau$ leptons}},  {\em Eur.
  Phys. J. C} {\bf 1} (1998) 221--233,
  [\href{http://arxiv.org/abs/hep-ph/9703344}{{\tt hep-ph/9703344}}].

\bibitem{Jezabek:1996ia}
M.~Jezabek and L.~Motyka, {\it {Perturbative QCD corrections to inclusive
  lepton distributions from semileptonic $b \to c \tau \bar \nu_\tau$ decays}},
   {\em Acta Phys. Polon. B} {\bf 27} (1996) 3603--3613,
  [\href{http://arxiv.org/abs/hep-ph/9609352}{{\tt hep-ph/9609352}}].

\bibitem{Czarnecki:1994bn}
A.~Czarnecki, M.~Jezabek, and J.~H. Kuhn, {\it {Radiative corrections to $b \to
  c \tau \bar \nu_\tau$}},  {\em Phys. Lett. B} {\bf 346} (1995) 335--341,
  [\href{http://arxiv.org/abs/hep-ph/9411282}{{\tt hep-ph/9411282}}].

\bibitem{Jezabek:1996db}
M.~Jezabek and L.~Motyka, {\it {Tau lepton distributions in semileptonic B
  decays}},  {\em Nucl. Phys. B} {\bf 501} (1997) 207--223,
  [\href{http://arxiv.org/abs/hep-ph/9701358}{{\tt hep-ph/9701358}}].

\bibitem{DeFazio:1999ptt}
F.~De~Fazio and M.~Neubert, {\it {$B \to X_u \ell \bar \nu_\ell$ decay
  distributions to order $\alpha_s$}},  {\em JHEP} {\bf 06} (1999) 017,
  [\href{http://arxiv.org/abs/hep-ph/9905351}{{\tt hep-ph/9905351}}].

\bibitem{Trott:2004xc}
M.~Trott, {\it {Improving extractions of $|V(cb)|$ and $m(b)$ from the hadronic
  invariant mass moments of semileptonic inclusive B decay}},  {\em Phys. Rev.
  D} {\bf 70} (2004) 073003, [\href{http://arxiv.org/abs/hep-ph/0402120}{{\tt
  hep-ph/0402120}}].

\bibitem{Aquila:2005hq}
V.~Aquila, P.~Gambino, G.~Ridolfi, and N.~Uraltsev, {\it {Perturbative
  corrections to semileptonic b decay distributions}},  {\em Nucl. Phys. B}
  {\bf 719} (2005) 77--102, [\href{http://arxiv.org/abs/hep-ph/0503083}{{\tt
  hep-ph/0503083}}].

\bibitem{Alberti:2014yda}
A.~Alberti et~al., {\it {Precision Determination of the
  Cabibbo-Kobayashi-Maskawa Element $V_{cb}$}},  {\em Phys. Rev. Lett.} {\bf
  114} (2015) 061802, [\href{http://arxiv.org/abs/1411.6560}{{\tt
  arXiv:1411.6560}}].

\bibitem{Colangelo:2019axi}
P.~Colangelo, F.~De~Fazio, and F.~Loparco, {\it {Probing New Physics with $\bar
  B \to \rho(770) \, \ell^- \bar \nu_\ell$ and $\bar B \to a_1(1260) \, \ell^-
  \bar \nu_\ell$}},  {\em Phys. Rev. D} {\bf 100} (2019) 075037,
  [\href{http://arxiv.org/abs/1906.07068}{{\tt arXiv:1906.07068}}].

\bibitem{Shi:2019gxi}
R.-X. Shi et~al., {\it {Revisiting the new-physics interpretation of the $b\to
  c\tau\nu$ data}},  {\em JHEP} {\bf 12} (2019) 065,
  [\href{http://arxiv.org/abs/1905.08498}{{\tt arXiv:1905.08498}}].

\bibitem{Zyla:2020}
{\bf Particle Data Group} Collaboration, P.~A. Zyla et~al., {\it {Review of
  Particle Physics}},  {\em Prog. Theor. Exp. Phys.} {\bf 2020} (2020) 083501.

\end{thebibliography}\endgroup

\end{document}